\documentclass{PoS}

\title{Flow, spectra and HBT radii in heavy-ion collisions}

\ShortTitle{Flow, spectra and HBT radii in heavy-ion collisions}

\author{\speaker{Piotr Bo\.zek}%
        \thanks{Supported by 
Polish Ministry of Science and Higher Education under
grant N202~034~32/0918}\\
       IFJ PAN, Krak\'ow, Poland\\
       E-mail: \email{piotr.bozek@ifj.edu.pl}}

\author{Iwona Wyskiel\\
         IFJ PAN, Krak\'ow, Poland}

\abstract{The 
expansion of the fireball created in 
 relativistic heavy ion collisions is described using the 
 $3+1$D hydrodynamical model. Experimentally observed
 transverse momentum spectra at 
different rapdities, 
elliptic flow and HBT correlations of produced particles can be reproduced.
We give estimates of shear viscosity corrections at freeze-out, 
which we find important only for the elliptic flow coefficient.}

\FullConference{European Physical Society Europhysics Conference on
High Energy Physics\\
                July 16-22, 2009\\
                Krakow, Poland}

\begin{document}

\begin{figure}
\begin{center}
\includegraphics[height=8cm]{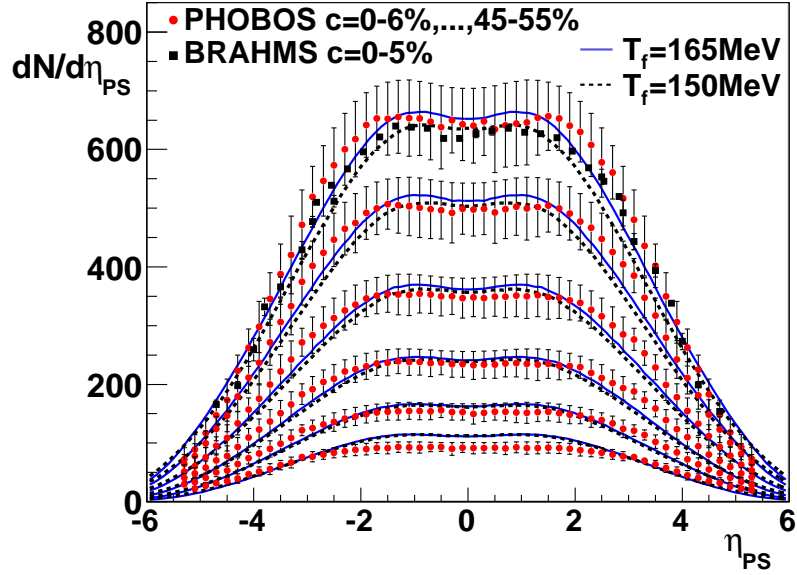}
\end{center}
\caption{Distribution of charged particles in pseudorapidity at 
different centralities, calculated from the 3+1D hydrodynamic 
model \cite{my}, compared to experimental data of
 BRAHMS and PHOBOS Collaborations.} \label{fig:dndy}
\end{figure}

The expansion of the dense and hot fireball created in relativistic heavy-ion collisions can be described using  relativistic hydrodynamics \cite{kolb}.
We perform   $3+1$D hydrodynamic simulations of Au-Au collisions at 
the highest RHIC energy  $\sqrt{s}=200$GeV \cite{my}. 
Compared to existing calculations \cite{Hirano:2002ds}, 
we use a very short initial time $0.25$fm/c and a realistic 
equation of state of dense matter \cite{Chojnacki:2007jc}. 
The absence of a soft point in the 
equation of state leads to a rapid expansion of the system, resulting in 
a strong build up of the transverse flow. 
The initial energy density profile in the transverse plane is 
taken from the Glauber Model, and the initial distribution in 
 space-time rapidity is adjusted to reproduce 
the measured charged particle distributions (Fig. \ref{fig:dndy}).

\begin{figure}
\includegraphics[height=14cm]{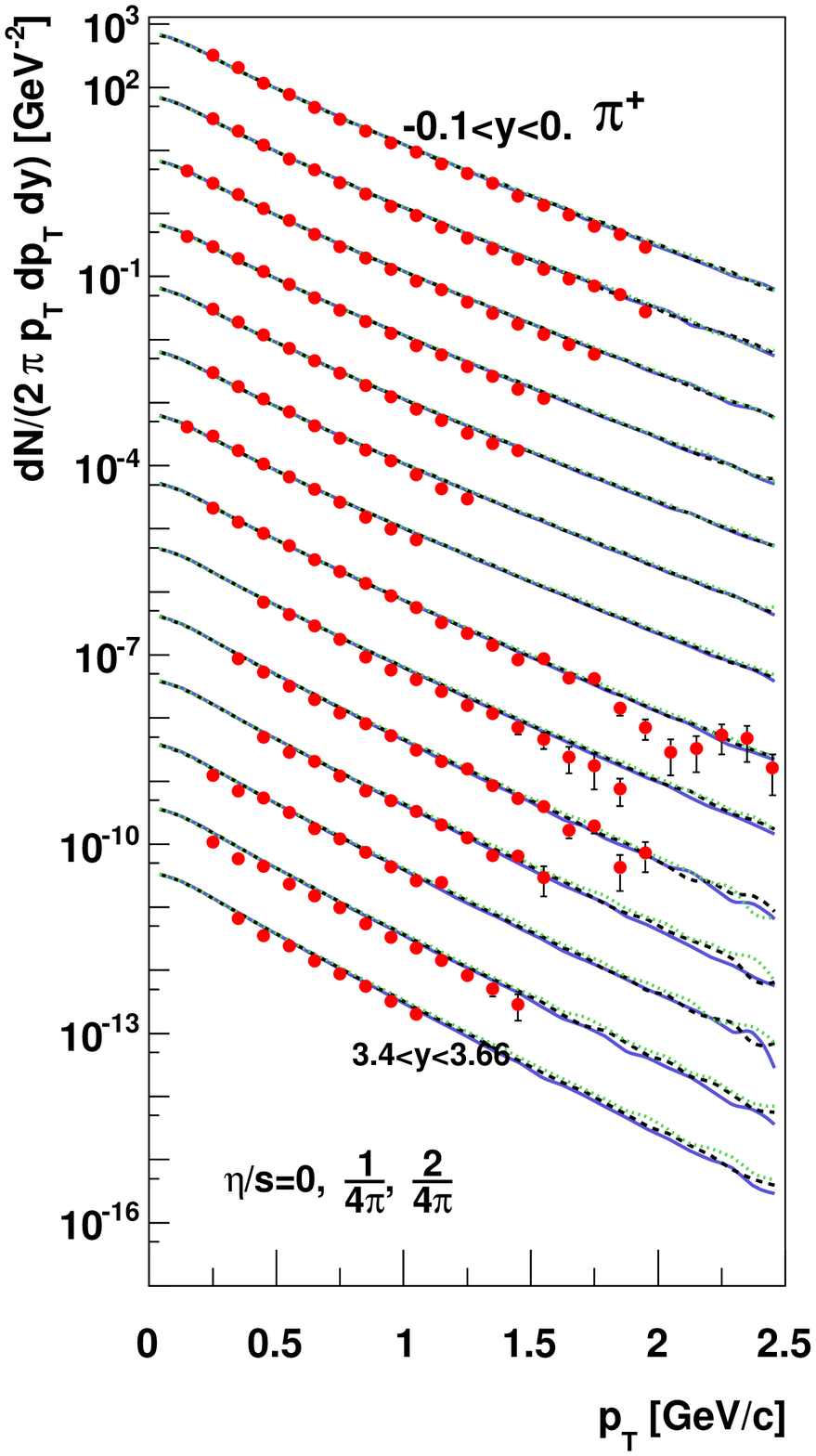}~~~\includegraphics[height=13cm]{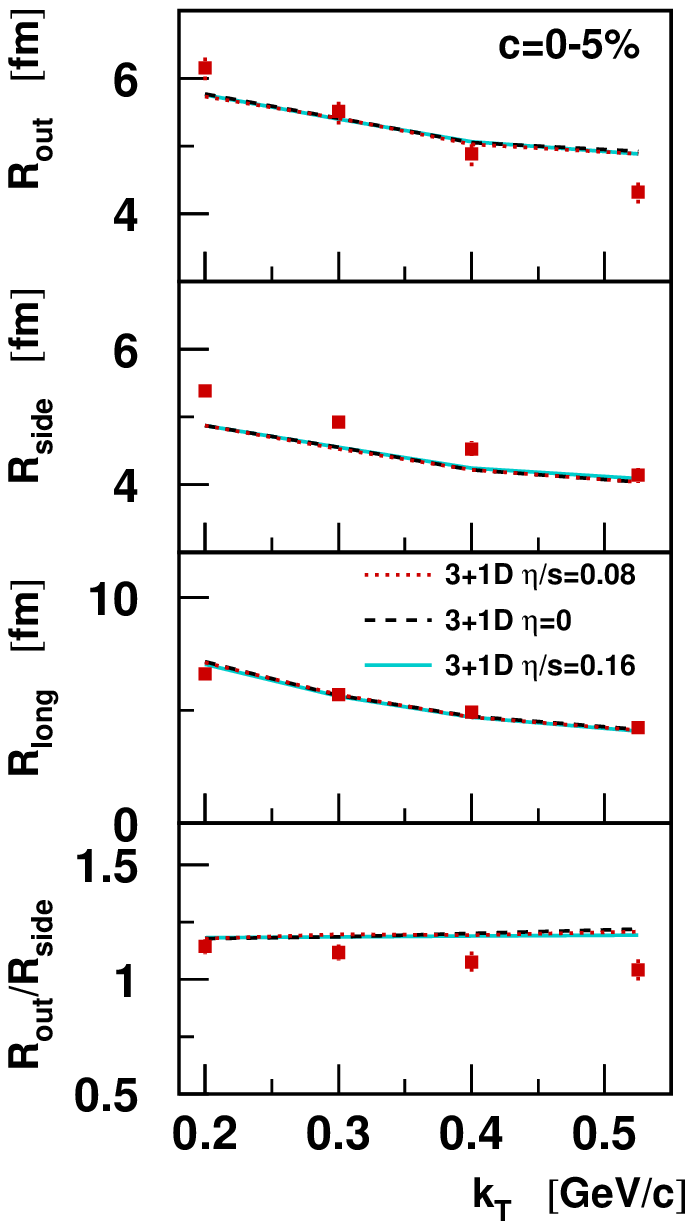}
\caption{(Left panel) Transverse momentum spectra of $\pi^+$ at different rapidities for three different shear viscosity coefficients at freeze-out ($\eta/s=0,\frac{1}{4\pi},\ \frac{1}{2\pi}$,  solid, dashed, and  dotted lines) compared to BRAHMS Collaboration data. (Right panel) HBT correlation radii for different strengths of  shear viscosity corrections
 at freeze-out, STAR Collaboration data.}
\label{fig:sp}
\end{figure}

\begin{figure}
\includegraphics[height=5.5cm]{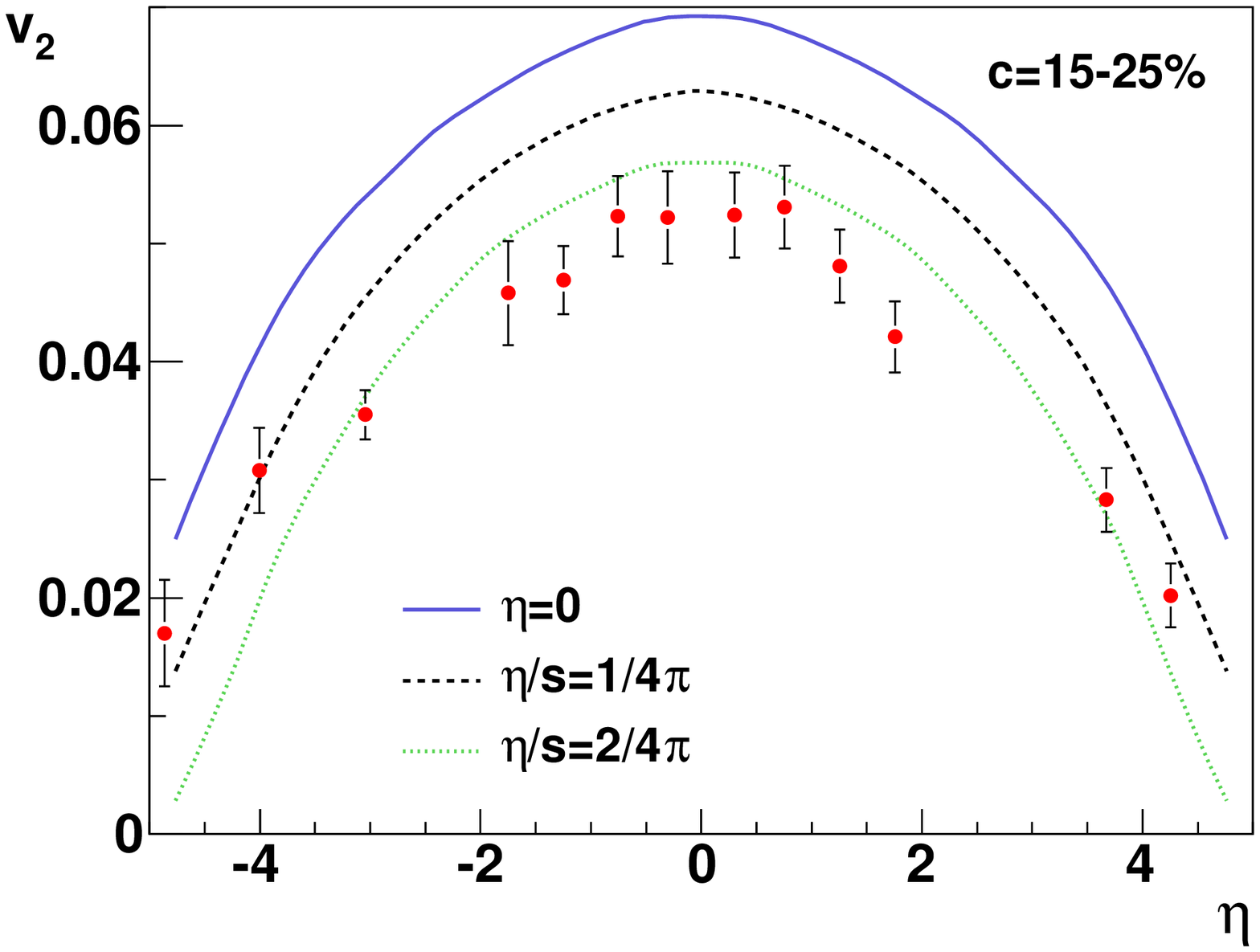}~\includegraphics[height=5cm]{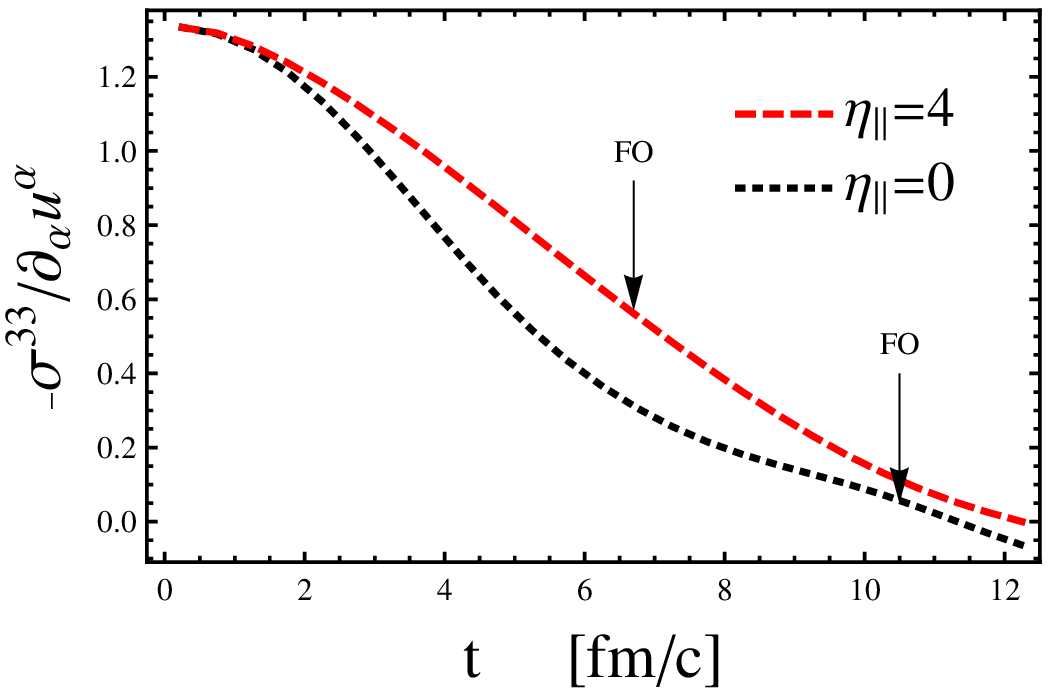}
\caption{(Left panel) Elliptic flow coefficient for charged particles as function of pseudorapidity for three different viscosity coefficients at freeze-out, compared to PHOBOS Collaboration data. (Right panel) Ratio
 of the stress velocity gradient to the overall expansion 
rate as function of time at two space-time rapidities. The 
arrows indicate the time of the freeze-out.}
\label{fig:v2}
\end{figure}

 Hydrodynamic equations of a 
perfect fluid  
$
\partial_\mu T^{\mu \nu}=0
$
are solved for each impact parameter. The shape of the constant temperature 
($T_F=150$MeV) freeze-out hypersurfaces,
 as well as the final collective
 velocities of the fluid are exported to a statistical emission and resonance 
decay code THERMINATOR \cite{Kisiel:2005hn}. The THERMINATOR code generates complete  events including statistically  emitted particles from the fireball.
 Calculated 
transverse momentum spectra of produced particles follow very well the 
experimental results \cite{my}. It is true for particle emitted at central 
rapidities
 for different collision centralities, up to $50\%$;
also the agreement with the data at 
forward/backward rapidities is striking (solid lines in the left panel of Fig. 
\ref{fig:sp}) and proves
 that the  hydrodynamic expansion model of the fireball combined 
with statistical  emission applies in a broad range of rapidities. 
Microscopically it means that approximate local equilibrium is  reached 
and maintained during a sizable time-span of the collective expansion. 
 HBT correlation radii \cite{Kisiel:2006is}  calculated from the model are
 within $10\%$ from 
the measured values (right panel of Fig. \ref{fig:sp}) \cite{my}.
A short initial time of the evolution requires a narrow initial distribution in 
space-time rapidity, this results in a strong suppression of elliptic flow when approaching  the fragmentation regions (Fig \ref{fig:v2}).

We estimate the effects of viscous corrections on  particle emission. 
Using the  velocity flow obtained in a $3+1$D  perfect fluid  
dynamics we calculate
  nonequilibrium corrections to the 
Cooper-Frye formula from shear viscosity \cite{Teaney:2003kp}.
Modified momentum distributions are implemented in the statistical
 emission code. Results range from, no appreciable modifications 
of the transverse momentum spectra for $p_\perp<1.5$GeV and central 
rapidities, up   to a $30\%$ increase at rapidity $y=3.5$
 and $p_\perp=3$GeV for $\eta/s=0.16$. The HBT radii are not sensitive at all
to the shear corrections at freeze-out if the the flow remains unchanged (Right panel in Fig. \ref{fig:sp}). 

A significant reduction of the elliptic flow is induced by 
 stress corrections (Fig. \ref{fig:v2}). The reduction is  $20\%$ at 
central rapidity and goes up $60\%$ at pseudorapidity $4$ for 
$\eta/s=0.16$. This observation agrees with the results of Ref. 
\cite{Hirano:2005xf}, where dissipation from hadronic rescattering 
was found to be important at large rapidities. This effect modifies strongly 
the dependence of the elliptic flow of charged particles on pseudorapidity.
From the ratio $\frac{-\sigma^{33}}{\partial_\alpha u^\alpha}= \frac{-2 \nabla^3 u^3 
+2/3\ \Delta^{33}\partial_\alpha u^\alpha}{\partial_\alpha u^\alpha}$
presented in the right panel of Fig. \ref{fig:v2} we 
expect large shear viscosity corrections at large rapidities, whereas 
at central rapidities correction from bulk viscosity 
(proportional to $\partial_\alpha u^\alpha$) could also be  important.
The approximate agreement of perfect fluid calculations with the 
data on elliptic flow
is accidental. Moreover shear viscosity  modifies also  the longitudinal
acceleration of matter  \cite{Bozek:2007qt} changing the distributions 
in Figs. \ref{fig:dndy} and \ref{fig:v2}.

Summarizing, we find good agreement of the results of $3+1$D
 perfect fluid hydrodynamics with the measured transverse momentum spectra,
 HBT radii and elliptic flow. Estimates of  shear viscosity effects show 
that the last observable is not robust and a reliable estimate thereof requires 
the use of  a fluid expansion model including shear and bulk viscosity.

\end{document}